# Enhancement of flux-line pinning in all-oxide superconductor/ferromagnet heterostructures

H.-U. Habermeier, J. Albrecht and S. Soltan
MPI-FKF, Stuttgart, Germany

**Abstract.** We have studied the local critical current density, $j_c$, in the superconductor thin film of bilayer structures consisting of $YBa_2Cu_3O_7$ and the ferromagnets $La_{2/3}Ca_{1/3}MnO_3$ and $SrRuO_3$, respectively, by means of quantitative magneto-optics. A pronounced hysteresis of $j_c$ was observed which is ascribed to the magnetization state of the ferromagnetic layer. The results are discussed within the frame of magnetic vortex – wall interactions.

## 1. Introduction

Vortex pinning in superconductors [SC] represents a major part in the field of vortex matter physics, and tailoring the interaction of vortices with pinning centers is a crucial prerequisite for the technological applications of SC's. Consequently, the exploration of mechanisms leading to vortex pinning and its artificial enhancement is a research topic which includes not only the interaction of the electronic structure of vortices with that at and around defects in the SC but also suprastrictive interactions of the elastic field generated by lattice defects with those of the SC/normal phase boundaries [1]. To address the pinning problem in high temperature superconductors [HTSC] where operating temperatures of 77K are envisaged is even more important due to their small coherence length, $\xi$, complexity of the magnetic field / temperature phase diagram and the role of thermal fluctuations which can cause substantial depinning. A general treatment of vortices in HTSC's is given by Blatter et.al. [2].

The majority of research work dealing with the pinning problem is focusing on the generation and study of microscopic defects as pinning centers by a variety of different methods. These range from processing in order to create dislocations and/or dislocation networks, heavy ion irradiation to generate columnar defects [3,4] to the manipulation of the growth mode of thin films to produce extended planar defects [5]. Additionally, methods have been employed where the pinning potential is artificially modulated by varying the SC film thickness [6] or the introduction of nanoscale antidot arrays [7]. These methods based on vortex - defect interaction have in common, that pinning forces result from inhomogeneities on the length scale of the coherence length. Disorder-induced spatial modulations of the linear and quadratic term of the Ginsburg-Landau functional $\alpha |\Psi|^2 + \beta/2 |\Psi|^4$ generate a suppression of the superconducting order parameter $\Psi$ and thus $T_C$ with the consequence that a defect causes the vortex to match its normal core with the region of reduced order parameter or finds a position to minimize its line energy. The upper limit for the pinning energy per unit length in this case corresponds to the Cooper pair condensation energy in the volume of the vortex core $U = (H_C^2/8\pi)\pi\xi^2 = (\Phi_0/8\pi\lambda_L)^2$ with $H_c$ being the thermodynamic critical field, $\Phi_0$ the flux quantum, and $\lambda_L$ the London penetration depth. Approaching $T_C$ this energy drops because of the increase of $\lambda_L$ and $\xi$, respectively. Additionally, disorder-induced spatial variations of the charge carrier mean free path, l, affect the nonlinear term $|grad\Psi|^2$ in the Ginsburg-Landau functional; these locations of enhanced carrier scattering facilitate the adjustment of the order parameter to the rapid changes required by the presence of the normal core of the vortex and cause pinning [8].

The alternative approach is to pin the magnetic flux rather than the vortex core has not been studied to that extend, an increasing number of papers, however, appeared in the past few years addressing this issue. In a rather early experiment Aoki and Habermeier [9] have determined the pinning force density per length due to ferromagnetic overlayers in plastically deformed Nb single crystals to 30 μN/m to 180 μN/m depending on the magnetic moment of material. The magnetic pinning was estimated to be substantially larger that that of the disclocation network. Helseth et. al [10] studied theoretically the interaction between a Bloch wall and a vortex considering the magnetostatic forces between a magnetic monopole (vortex) and a magnetic surface change (Bloch wall). They determined pinning forces in the range of (1–1.5) x$10^{-10}$N depending on the magnetic moment. Bulaevskii et al [11] proposed magnetic flux pinning in nanoscale SC- ferromagnet superlattices and estimated the maximum magnetic pinning energy for a single vortex to be 100 times larger than the pinning energy of columnar defects. Flux pinning by magnetic dipoles and regular arrays of nanoscale magnetic dots in conventional SC thin films have been studied by the groups of Schuller [12,13] and Bruynseraede [14] and there is consensus that the magnetic interaction plays an important role in the matching effects observed.

In this paper previously reported hysteretic effects [15,16] observed magneto-optically in YBCO / ferromagnetic oxide heterostructures are presented and compared within the frame of the above-mentioned magnetic pinning mechanisms.

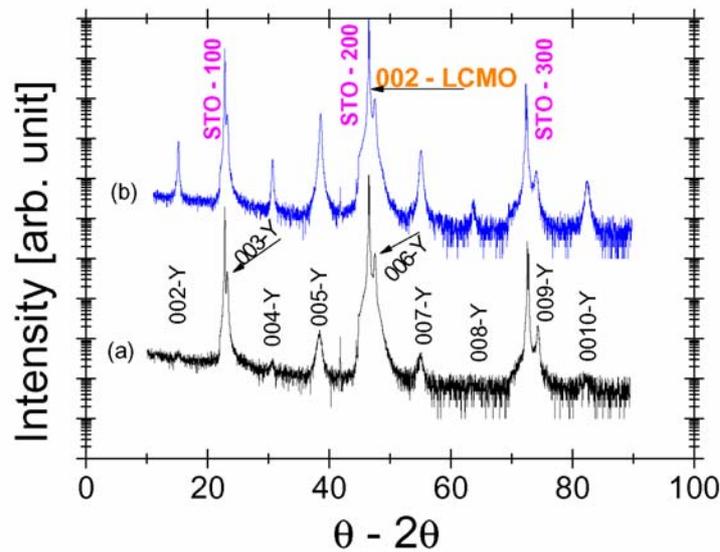

Fig. 1  X-ray diffraction pattern of a bilayer with $t_{LCMO}$ = 50 nm / $t_{YBCO}$ = 20 nm (a) and $t_{LCMO}$ = 50 nm / $t_{YBCO}$ = 100 nm (b)

## 2. Specimen preparation and characterization

The ferromagnetic [FM] materials used for this study, $La_{2/3}Ca_{1/3}MnO_3$ [LCMO] and $SrRuO_3$ [SRO], have been selected because of their compatibility in the deposition process with that of $YBa_2Cu_3O_7$ [YBCO], their good lattice match with YBCO [0.4 % YBCO/LCMO and 2% YBCO/SRO], their different saturation magnetization [3.5$\mu_B$ for LCMO and 1.5$\mu_B$ for SRO] and different mechanisms giving rise to ferromagnetism. In LCMO FM order is caused by the Zener double exchange mechanism whereas SRO is an itinerant band ferromagnet with strong electron correlation. The bilayer structures have been grown on $SrTiO_3$ single crystal substrates with (100) orientation by standard pulsed laser deposition at $770^0C$ and an oxygen partial pressure of 0.5 mbar. The individual layer thickness ranged from 20nm to 100 nm each, controlled by computerized pulse counting during the deposition process. The FM layers are used as a functional buffer for the YBCO film. A more detailed description of the deposition process is given in [17]. The layers are single phase and epitaxially grown as revealed by X-ray diffraction (c.f. Fig. 1) and TEM analysis [18,19]. Transport and magnetization measurements complemented the structural analysis to determine the Curietemperature, $T_{Curie}$, and $T_C$ as well. In analogy to results obtained in YBCO/LCMO and YBCO/SRO superlattices [19,20] there is a reduction of $T_{Curie}$ and $T_C$ depending on the ratio and individual layer thickness of the FM and SC part of the bilayer. Fig. 2 gives two representative examples for the YBCO/LCMO heterostructures for 30 nm YBCO / 50 nm LCMO/ and 100 nm YBCO / 50 nm LCMO, respectively. Complementary ellipsometry data [20] as well as transport and magnetic investigations in FM/SC superlattices suggest that these depressions are based on an electronic interaction between the FM and SC part rather than on extrinsic effects such as lattice strain, interdiffusion, incomplete oxygenation etc. [17-21]. This electronic interaction causes a saturation magnetization dependent reduction of the critical current density, $j_c$, in the film up to one order of magnitude. Replacing the FM layer by a nonmagnetic metallic oxide layer such as $LaNiO_3$, restores $j_c$ to values close to those typical for single layer YBCO. This evidences that $T_C$ and $j_c$ reduction is not an extrinsic effect.

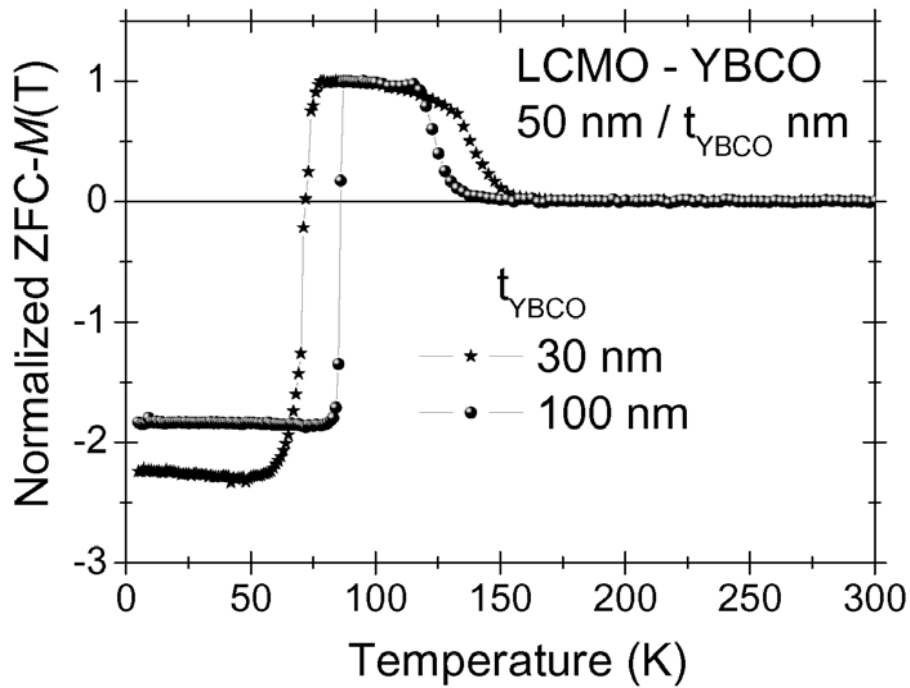

Fig.2  Temperature dependence of the normalized magnetization [$M(T)/M_{300}$] for a 50nm/30nm LCMO/YBCO and a 50nm LCMO/100nm YBCO heterostructure, respectively, measured in an in-plane field of 10 Oe.

By means of quantitative magneto-optics, the magnetic flux distribution after applying different magnetic fields perpendicular to the film plane was measured at 7K. The technique makes use of the magneto-optical Faraday effect in FeGdY garnet indicator films. The determination of the local critical current is based on a naumerical inversion scheme of Biot-Savart's law. Here, the measured z-component of the magnetic flux is related to the in plane current in the sample. This technique enables the determination of the corresponding critical current distribution with a spatial resolution of ~ 5 µm as described by Jooss et.al [22].

## 3. Experimental Results
### 3.1 YBCO-LCMO Heterostructures

Previous magneto-optical as well as transport current experiments to determine $j_c$ in single layer YBCO films deposited on $SrTiO_3$ single crystals revealed a homogeneous $j_c$ distribution with typical values (at 5 K) of $j_c \sim 3 \times 10^{11}$ $A/m^2$ [22]. This value is regarded as a standard for high quality YBCO thin films in the literature. In Fig. 3a,b the magneto-optically determined current density distribution of a 50nm YBCO/50 nm LCMO heterostructure is shown in an external magnetic field of $\mu_0 H_{ex} = 3mT$ applied perpendicular to the film plane (T = 7K). The image in Fig. 3a is taken after field cooling whereas Fig.3b represents the current distribution in decreasing field after applying $\mu_0 H_{ex} = 100mT$. The grey scale representation refers to a linear increase of the current density from $5 \times 10^9$ $A/m^2$ (black) to $6 \times 10^{10}$ $A/m^2$ (white). Qualitatively it is obvious that in

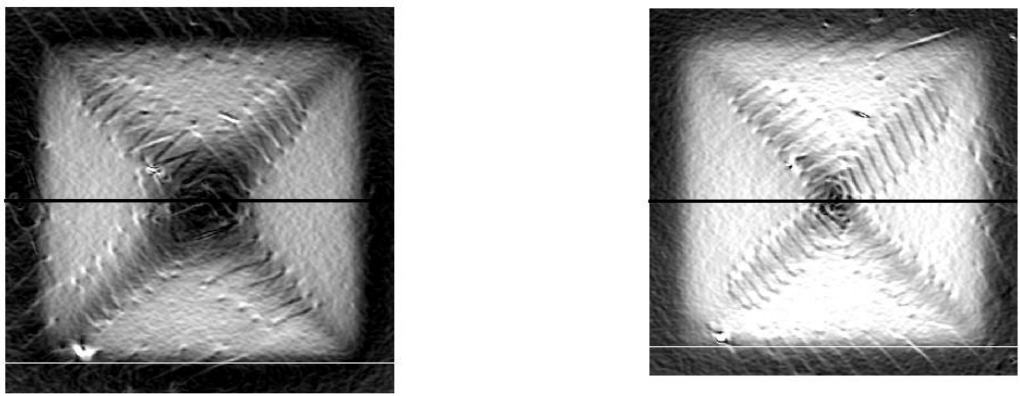

Fig.3 Current density in the 50nmYBCO/50nmLCMO heterostructures [5 x 5 $mm^2$] at an external field of $\mu_0 H_{ex} = 3mT$ after zero field cooling (Fig 3a left ) and in decreasing field from 100 mT ( Fig.3b right ).

Fig. 3b white dominates with an average current density of $j_c \geq 6 \times 10^{10}$ A/m$^2$ whereas Fig. 3a remains more dark with $j_c$ around $4 \times 10^{10}$ A/m$^2$ (only the part with penetrated flux has to be considered). A line scan indicated in Fig. 3 enables a more quantitative comparison of the data as shown in Fig. 4. The lower profile in Fig. 4 corresponds to the zero field cooled measurement whereas the upper one is determined after the application of $\mu_0 H_{ex}$ = 100mT and then reduced to 3mT. The averaged enhancement $\Delta j_c = 1.5 \times 10^{10}$ A/m$^2$ - corresponding to an additional pinning force density per length of 30 µN/m - is due to the magnetic history of the ferromagnetic layer. Qualitatively, the enhancement of $j_c$ is believed to be associated with the magnetic domain structure and its magnetic stray field interacting with that of the flux-line lattice. Consequently, the contributions of magnetic pinning in the bilayer structures should vary with the relative size of the interaction volume of flux lines and magnetic closure structure, i.e. the effect is expected to be dependent on the YBCO film thickness. Indeed, magneto-optical measurements of $j_c$ in 100nmYBCO/50nmLCMO bilayers reveal $\Delta j_c = 0.8 \times 10^{10}$ A/m$^2$, just about the half of the values for the 50nmYBCO/50nmLCMO bilayers. It is important to note that the critical current density itself depends on the layer thickness of YBCO. Whereas the 50nmYBCO/50nmLCMO structure has $j_c \sim 4 \times 10^{10}$ A/m$^2$ the 100nmYBCO/50nmLCMO has $j_c \sim 8 \times 10^{10}$ A/m$^2$.

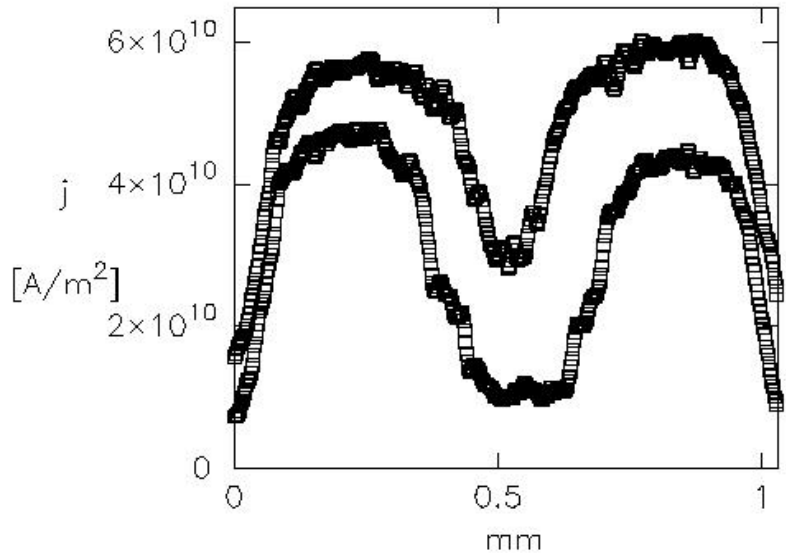

Fig. 4  Averaged current density profiles along the horizontal lines in Fig.3. The lower curve refers to the zero field cooling measurement.

### 3.2   YBCO-SRO Heterostructures

Complementary to the experiments using the colossal magnetoresistance material LCMO we have studied similar samples with SrRuO$_3$ as ferromagnetic layer. The main results of the measurements performed on 100nmYBCO/50nmSRO bilayers can be summarized as follows:

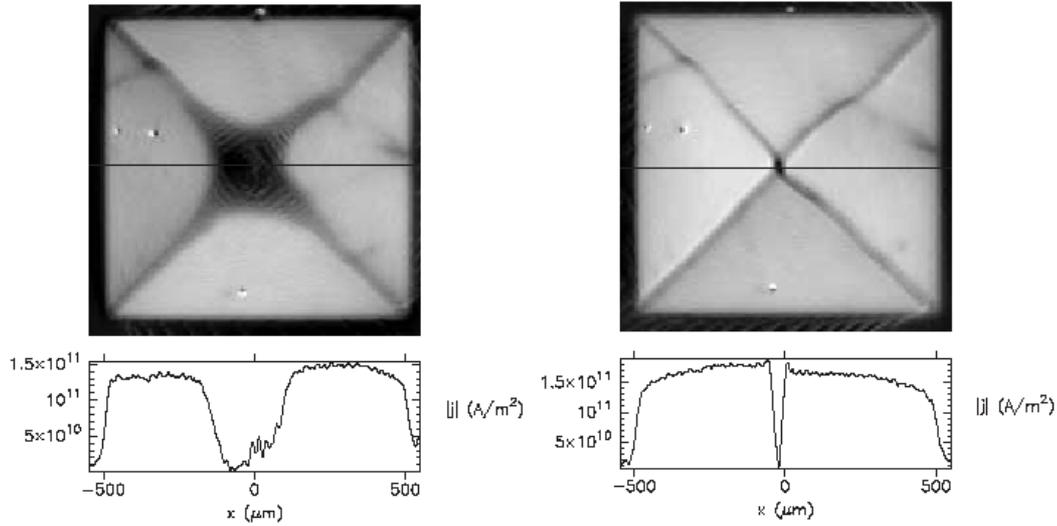

Fig. 5 Grey scale representation of the critical current density and the corresponding profiles along the solid black lines

1.  As shown in the gray scale representation of the magnitude of the critical current distribution and the corresponding profiles along the solid black lines in Fig. 5 there is a clear difference in the averaged values for $j_c$ measured at 16 mT in the case of zero field cooled samples (left image) and in 16 mT after applying an external field of 200 mT. A quantitative analysis of the averaged values reveals an enhancement of $\Delta j_c = 3 \times 10^{10}$ A/m$^2$ with an average value for $j_c = 1.6 \times 10^{11}$ A/m$^2$ for the measurement after zero field cooling. A more detailed description of the evaluation procedures taking into account the existence of in homogeneities of the sample is given in [16].

2.  The magnetic field dependence of the critical current is given in Fig. 6. The lower branch represents $j_c$ in increasing field after zero field cooling, the upper one $j_c$ after applying $\mu_0 H_{ex} = 200$ mT and then gradually reducing the field. The maximum field applied for the measurements was confined to $\mu_0 H_{ex} < 50$ mT to avoid saturation effects in the magneto-optical indicator film. There is a clear hysteresis effect, which is due to the magnetization state of the magnetic buffer layer. All measured $j_c$ values after applying $\mu_0 H_{ex} = 200$ mT are well above those measured after zero field cooling. Quantitatively, the enhancement of the critical current corresponds to an additional pinning force density per length of approximately 60 µN/m.

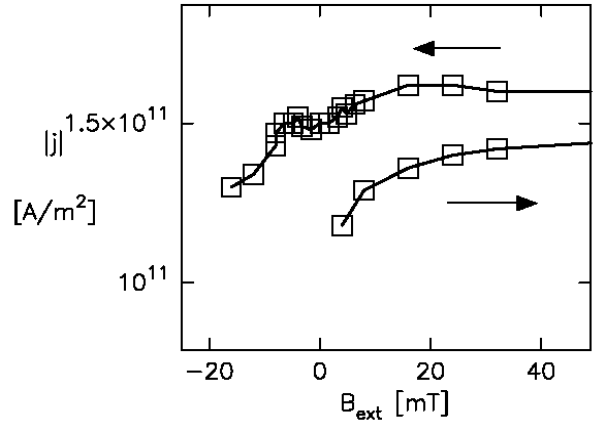

Fig.6 Hysteresis of the critical current density in the YBCO film. The lower branch corresponds to rising field after zero field cooling, the upper to a gradually reduced field after applying $\mu_0 H_{ex} = 200 mT$.

3. In both branches of the magnetic field dependence of $j_c$ an increase with the applied field is observed.

4. **Discussion**

Qualitatively, it is demonstrated, that the magnetic buffer layer causes hysteresis in $j_c$ and an enhancement of $j_c$ after magnetizing the LCMO or SRO film perpendicular to the film plane. The reference sample YBCO/LaNiO does not show the hysteresis effect and displays the features of a single layer YBCO film. Furthermore there is an increase in $j_c$ after zero field cooling with increasing applied field contrasting to the well known usual decrease of $j_c$ in a magnetic field. This holds especially, if $\mu_0 H_{ex}$ is much higher that the self-field generated by the currents in the sample. This is not necessarily valid for small values for $\mu_0 H_{ex}$ usually applied in magneto-optics where local current densities are measured. In a first approximation there is a $j_c \sim \sqrt{B}$ dependence, which can be described by a model using the theory of elasticity of flux-lines assuming a lattice of heavily, deformed vortices [8].

To shed some light on the physics of this interaction some comments on the magnetic behavior of the ferromagnetic buffer layer is necessary. Generally, minimizing the magnetostatic energy, thin FM films have an in-plane orientation of the magnetization vector; in-plane tensile epitaxial strain as in the case of the STO/LCMO as well as STO/SRO interface causes a biaxial in-plane magnetic anisotropy. In a magnetic field applied perpendicular to the film plane, the magnetization process is dominated by the rotation of the magnetization vector out of plane rather than by wall movements thus causing an S shaped hysteresisloop. Square shaped hysteresisloops are expected if the external field is applied parallel to the easy axis of the magnetic film. The existence of a hysteresisloop is a clear hint that domain wall movements and domain wall pinning play a role in the magnetization process, pure magnetization rotation would be hysteresis free. Two different scenarios have to be distinguished, perpendicular magnetic anisotropy and in-plane biaxial magnetic anisotropy. In the case of perpendicular magnetic anisotropy and the presence of stripe domains the model of Bulaevskii can be applied [9]. The stripe domains cause a change of sign effective magnetic field B (x) = $\mu_0 H_{ex}$ +4M(x) in the superconductor thus creating a pinning potential for vortices. According to this model the magnetic pinning should scale with the saturation magnetization M and the maximum value for the pinning potential can be estimated from $U_{mp} \sim \Phi_0 M_S d_S$ where $M_S$ is the saturation magnetization and $d_S$ the thickness of the superconductor film. An experimental verification of this can be seen in the measured shift of the irreversibility line in YBCO / Ba-orthoferrite heterostructures where the magnetic layer has a perpendicular magnetic anisotropy [23]. In the case of in-plane magnetic anisotropy the stray fields of domain walls causes the magnetic field modulation in the superconductor. Therefore a pinning effect due to vortex/domain wall interactions is expected to occur in the field range where domain reorganization plays a role, i.e. below the coercive field.

| Sample Type | 50nmYBCO/ 50nmLCMO | 100nmYBCO/ 50nmLCMO | 100nmYBCO/ 50nmSRO |
|---|---|---|---|
| $T_C$ [K] | 84.7 | 86.7 | 87.9 |
| $j_c$ [$10^{10}$ A/m$^2$] | 4 | 8 | 16 |
| $\Delta j_c$ [$10^{10}$ A/m$^2$] | 1.5 | 0.8 | 3 |
| $\Delta j_c / j_c$ | 0.375 | 0.1 | 0.18 |

Tab. I  Comparison of the critical current enhancement for the different sample types

A quantitative comparison of the $j_c$ enhancement for the three different types of samples is given in Tab.I. The largest relative enhancement of $j_c$ has been found in the 50nmYBCO/50nmLCMO sample the value drops drastically for the 100nmYBCO / 50nmLCMO sample. This clearly indicates that the origin of the $j_c$ enhancement arises from the interface of both layers. A comparison of the $j_c$ enhancement for the 100nmYBCO/50nmLCMO and the corresponding structure with the SRO shows a surprisingly larger effect for the magnetic layer with the smaller saturation magnetization. Ruling out that extrinsic effects such as lattice strain, interdiffusion, incomplete oxygenation or massive charge transfer can be accounted for the unexpected differences in the material dependence of $\Delta j_c$, the experimental results can only be explained with the different domain structures of the layers involved. Furthermore, the thickness of the magnetic layers (50 nm) is in a range where the transition from Blochwalls to Nèelwalls is expected to take place and their quite different magnetic stray field configurations will cause different pinning energies. A quantitative understanding of the hysteresis effects in $j_c$ in the bilayers requires not only the study of hysteresis curves of single layer LCMO and SRO thin films in conjunction with domain structure analysis but also their modification in the presence of a vortex lattice in the adjacent layer. As already pointed out by Helseth [10] the presence of a superconductor reduces the equilibrium width of a Bloch wall, the determination of the magnetic domain structure in the presence of a lattice of magnetic monopoles (vortices) with a magnetic field dependent lattice constant at the surface remains an interesting task.

In summary, we have demonstrated a substantial enhancement of $j_c$ in superconductor / ferromagnet bilayers. The observed hysteresis in $j_c$ is attributed to the hysteresis of the magnetic buffer layers. In single layer YBCO films such effects never have been observed; this rules out influences due to sample roughness and hysteresis effects due to the order/disorder phase transition in the temperature/magnetic field phase diagram. Further investigations will concentrate on systems where the magnetic layer has a perpendicular anisotropy and is electronically decoupled from the superconductor.

**References**

[1]     Campbell A. and Evetts J. 1972 *Adv. Phys.* **21** 199

[2]     Blatter, G, Feigel'man, M. V., Geshkenbei, V. B., Larkin, A. I. And Vinokur, V. M. 1994 *Rev. of Mod. Phys.* **66**, 1125

[3]     Civale L, Marwick A D, Worthington T K, Kirk M A,

[4]     Nelson D R and Vinokur V M 1992 *Phys. Rev. Lett.* **68** 2398

[5]     Haage T, Zegenhagen J, Li J Q, Habermeier H U, Cardona M, Jooss Ch, Warthmann R, Forkl A and Kronmüller H 1997 Phys. Rev. B **56**, 8404

[6]     Daldini O, Olsen J L and Berner G 1974 *Phys. Rev. Lett.* **32** 218

[7]     Baert M, Metlushko V V, Jonckheere R, Moshalkov V V and Bruynseraede Y 1995 *Phys. Rev.Lett.* **74** 3269

[8]     Brandt E H 1995 *Rep. Prog. Phys.* **58** 1465

[9]     Aoki R and Habermeier H U 1987 *Jap. J. Appl. Phys.* **26** 1453

[10]    Helseth L E , Goa P E , Hauglin H, Baziljevich M, and Johansen T H 2002 *Phys. Rev. B.* **65** 132514

[11]    Bulaevskii L N, Chudnovski E M and Maley M P 2000 *Appl. Phys. Lett.* **76** 2594

[12]    Hoffmann A, Prieto P and Schuller I K 2000 *Phys. Rev. B* **61** 6958

[13]    Stoll O M, Montero M I, Guimpel J, Akerman J J and Schuller I K 2002 *Phys. Rev. B* **65** 104518

[14]    Harada K, Kamimura H,Kasai T, Matsuda A, Tonomura A and Moshalkov V V 1996 **Science** 274 1167

[15]    Albrecht J, Soltan S and Habermeier H-U 2003 *Physica C* in the print

[16]    Albrecht J, Soltan S and Habermeier H-U 2003 *Europhys. Lett.* **63** 881

[17]    Habermeier H U 1991 *Eur. J. Solid State Inorg. Chem* **28**

[18]    Habermeier H U and Cristiani G, 2003 *IEEE Trans. on Appl. Supercond.* **13** 2842

[19]    Habermeier H U et, al., 2001 *Physica C* **364-365** 298

[20]    Holden T, Habermeier H U, Cristiani G, Golnik A, Boris, Pimenov A , Humlicek J, Lebedev O, Van Tendeloo G, Keimer B, and Bernhard C, . B 2003 – cond. Mat./0303284.

[21]    Habermeier H U and Cristiani G 2002 , *Proceedings of SPIE* **4811** 111 (2002).

[22]    Jooss Ch, Albrecht J, Kuhn H, Leonhasrdt S and Krinmüller H 2002 *Rep. Prog. Phys.* **65** 651

[23]    Garcia-Santiago A, Sanchez F, Varela M and Tejada J 2000 *Appl. Phys. Lett.* **77** 2900